\documentclass[reprint,prl,twocolumn]{revtex4-2} 
\usepackage{amsmath}
\usepackage{amsfonts}
\usepackage{graphicx}
\usepackage{color}
\usepackage{xcolor}
\usepackage{float}
\usepackage{soul}

\begin{document}
\title {Theory of high-power excitation spectra of rf-SQUID}
\author{Olesia Dmytruk, R.~H.~Rodriguez, {\c{C}}.~{\"O}.~Girit, and Marco Schir\'o}
\affiliation{JEIP, USR 3573 CNRS, Coll\`ege de France, PSL Research University, F-75321 Paris, France}
\date{\today}

\begin{abstract}
We discuss the theory of linear and non-linear spectroscopy of an rf-SQUID coupled to a Josephson spectrometer.  Recent experimental measurements on this system have shown a strongly non-linear absorption lineshape, whose current peak maximum undergoes a forward-backward bending transition depending on the value of the rf-SQUID phase. We show that this transition can be qualitatively understood by mapping the dynamics of the driven rf-SQUID onto a generalized Duffing oscillator, with tunable drive and non-linearity, undergoing a bifurcation. Finally we show that in order to quantitatively reproduce the experimental data reported in~\textit{arXiv:2106.02632}, it is crucial to include the feedback from the load-line, leading to an additional source of non-linearity.

\end{abstract}

\maketitle

{\it Introduction.}~---~ 
The setups based on Josephson junctions have been at the center of research attention for many years.
Such setups are very versatile, as they can be used as  qubits~\cite{makhlin1999josephson,makhlin2001quantum,vion2002manipulating,wallraff2004strong,wendin2007quantum,ladd2010quantum,buluta2011natural,wendin2017quantum}, metamaterials~\cite{jung2014multistability}, Josephson bifurcation amplifiers~\cite{siddiqi2004rf,siddiqi2005direct,boutin2021topological}   
or detectors of mesoscopic systems~\cite{edstam1994josephson,holst1994effect,lindell2003quantum,billangeon2007very,petkovic2009direct,basset2012high,bretheau2013exciting,van2017microwave}. Moreover, Josephson junctions with external time-dependent driving are suitable platforms for studying nonlinear phenomena~\cite{dykman2012fluctuating,manucharyan2007microwave,zorin2011period,divincenzo2012nonlinear,gosner2019quantum,lang2021multi}.

Absorption spectroscopy of Josephson junction is a powerful experimental technique that can be used to study mesoscopic systems in a wide frequency range~\cite{bretheau2013exciting,bretheau2014theory}. Very recently a novel Josephson junction spectrometer with broad bandwidth and variable coupling strength was implemented and used to perform high-power spectroscopy on an rf-SQUID~\cite{griesmar2021superconducting}. The current-voltage characteristic of the spectrometer, related to the system absorption, was found to depend strongly on the phase of the rf-SQUID, $\varphi_x$. In particular  the position of the current maximum was found to shift towards higher or lower frequencies depending on $\varphi_x$, resulting in a forward or backward bending of the absorption peak~\cite{griesmar2021superconducting}.

\begin{figure}[t]
	\centering
	\includegraphics[width=\linewidth]{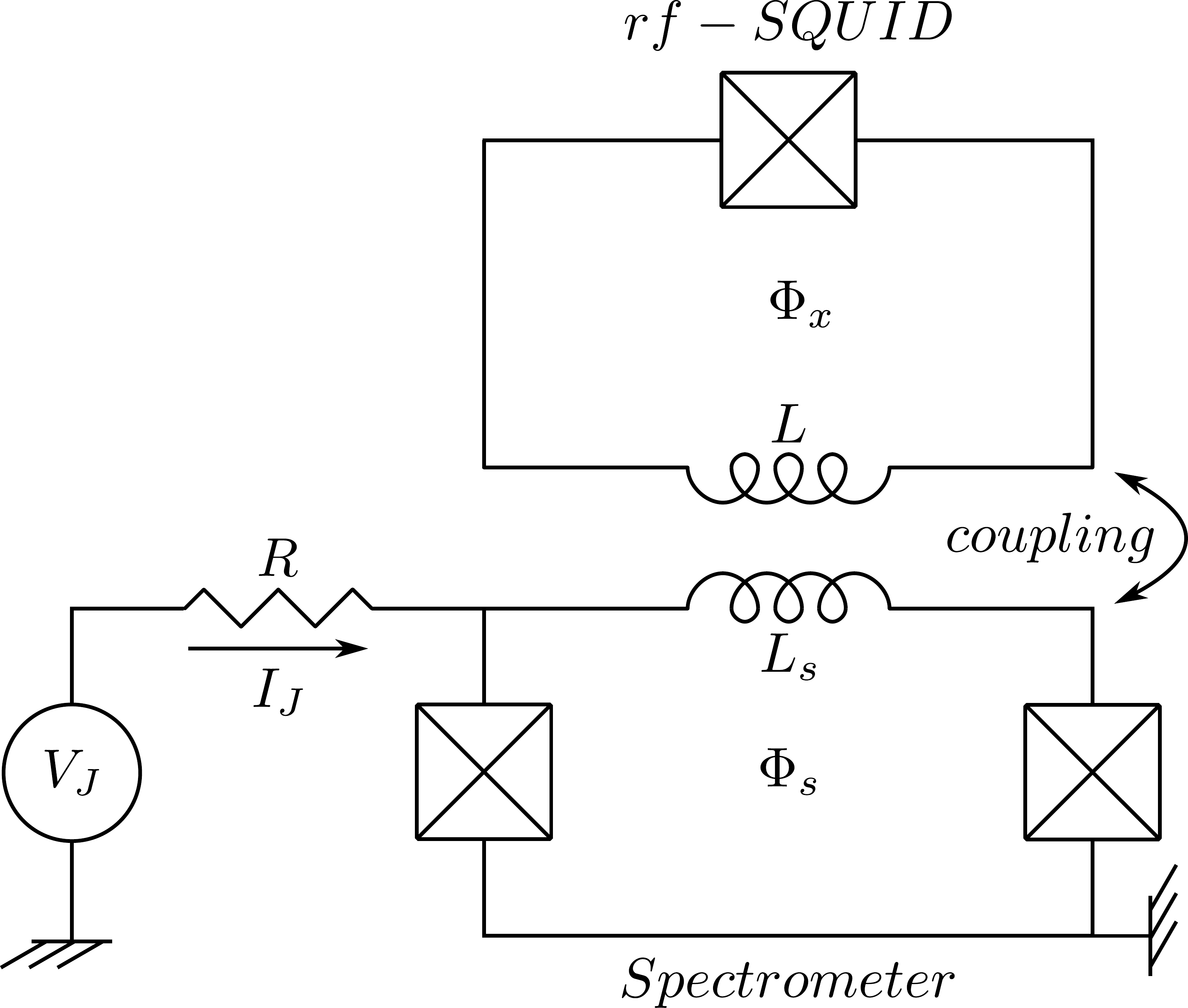}
	\caption{Scheme of the setup: an rf-SQUID (top) coupled inductively to a Josephson junction spectrometer (bottom). The rf-SQUID is formed by a  superconducting loop, threaded by a magnetic flux $\Phi_x$, of inductance $L$ with a single Josephson junction (boxed cross).
The Josephson spectrometer consists of two voltage-biased Josephson junctions  (boxed crosses) in a superconducting loop of inductance $L_s$ enclosing a static magnetic flux $\Phi_s$. 
	Bias circuit with voltage source $V_J$ and resistor $R$ supplies dc current $I_J$.
	}
		\label{fig:setup}
\end{figure}

Motivated by this experiment, in this Letter we present an effective model for an rf-SQUID strongly driven by the Josephson junction spectrometer and discuss its linear and non-linear spectroscopy. The rf-SQUID consists of a single Josephson junction in a superconducting loop enclosing a magnetic flux $\Phi_x$ and it is inductively coupled to the spectrometer, which is formed by two Josephson junctions in a superconducting loop threaded by a magnetic flux $\Phi_s$,  see Fig.~\ref{fig:setup}. In our effective model this inductive coupling results in a periodic driving of the rf-SQUID at a frequency set by the biasing voltage and amplitude controlled by $\Phi_s$.  We map the dynamics of the driven rf-SQUID in the semi-classical regime onto a generalized Duffing oscillator with tunable parameters. In particular we show that the sign of the leading Duffing non-linearity can be tuned by changing the phase of the rf-SQUID, resulting in a forward-backward bending transition of the absorption spectrum.  While capturing the qualitative features of the experiment, the mapping to the Duffing oscillator cannot by itself reproduce the observed lineshape, which features a strong asymmetry between backward and forward bending. We show that accounting for the feedback from the load-line, another key feature of the setup of Ref.~\cite{griesmar2021superconducting}, leads to an additional and sizable source of non-linearity which is crucial to quantitatively reproduce the experimental results.

{\it Effective Model for rf-SQUID coupled to Spectrometer.}~---~ 
To model the setup in Fig.~\ref{fig:setup} we consider an rf-SQUID whose Hamiltonian reads~\cite{makhlin2001quantum} 

\begin{align}
	H_{rf}  = E_C \hat{N}^2+ E_L [\hat{\varphi} - \varphi_{x}]^2 	- E_J \cos(\hat{\varphi})\,.
	\label{eq:HamiltonianRF}
\end{align}
Here, the first term describes the charging energy $E_C=2 e^2/C$, with $C$ the capacitance of the junction and $\hat{N}$ the number of Cooper pairs conjugated to the phase $\hat{\varphi}$, $[\hat{N},\hat{\varphi}] = -i$, while the second term accounts for the inductive energy $E_L =\varphi_0^2/\left(2L\right)$, with $L$ the self-inductance of the loop, $\varphi_x = \Phi_x/\varphi_0$ the phase of the rf-SQUID, $\Phi_x$ the magnetic flux threading the loop and $\varphi_0 = \Phi_0/\left(2\pi\right)$ the reduced flux quantum. Finally, the last term in Eq.~\ref{eq:HamiltonianRF} describes the Josephson non-linearity of strength $E_J = I_0 \varphi_0$, with $I_0$ the critical current.

A microscopic model of the coupling between rf-SQUID and spectrometer, including the basic quantum degrees of freedom of the latter is discussed in detail in Ref.~\footnote{See Supplemental Material at [URL will be inserted by publisher] for the derivation of the microscopic Hamiltonian starting from the full circuit}. Here we present an effective description according to which the inductive coupling between rf-SQUID and spectrometer leads, in presence of a finite voltage $V_J$ biasing the latter, to an explicit periodic driving for the former at the Josephson frequency $\omega_J = 2 e V_J/\hbar$ i.e.

\begin{align}
	H_{coupl} =  -2 E_L A(\varphi_s)\cos(\omega_J t )\hat{\varphi},
\label{eq:HamiltonianCoupling}
\end{align}
where 
$A(\varphi_s) = k I_{os}\varphi_0\sqrt{L_s/L}/\left(2E_L\right) \sin\left(\varphi_s/2\right)$.  Here, $k$ is the coupling coefficient between two inductive loops resulting from their mutual inductance,  $L_s$ is the inductance of the spectrometer loop, $\varphi_s = \Phi_s/\varphi_0$ is the phase difference across the spectrometer, with $\Phi_s$ being the magnetic flux through the spectrometer loop.

In order to include the dissipation, we couple the rf-SQUID to the bosonic bath, $H_{bath} = \sum_{\alpha}\hbar\omega_\alpha \hat{b}^\dag_{\alpha} \hat{b}_{\alpha}$ such that we get
\begin{align}
	H = H_{rf} + H_{coupl} +H_{bath}+ \hat{\varphi}\sum_{\alpha} g_\alpha(\hat{b}_\alpha + \hat{b}^\dag_\alpha),
	\label{eq:HamiltonianTotal}
\end{align}
where $\hat{b}_\alpha^\dag$ ($\hat{b}_\alpha$) are bosonic creation (annihilation) operators of the bath, $\omega_\alpha$ is the frequency of the bosonic bath, $g_\alpha$ is the coupling strength between the bosonic bath and rf-SQUID and we assume an Ohmic spectral function for the bath.

{\it Semiclassical Dynamics of Driven-Dissipative rf-SQUID.}~---~ In this paper we focus on the regime $E_L\gg E_J\gg E_C$, which is relevant for the setup of Ref.~\cite{griesmar2021superconducting}.  For $E_J \gg E_C$ the flux $\hat{\varphi}$ is the quantum degree of freedom~\cite{makhlin2001quantum}. Therefore, we define~\cite{bretheau2014theory} $\hat{\varphi} =  \sqrt{2}\kappa \hat{X}$ and $\hat{N} = \hat{P}/\left(\sqrt{2}\kappa\right)$, where
$\hat{X},\hat{P}$ are harmonic oscillator variables, related to the bosonic creation (annihilation) operators of the plasma mode of the rf-SQUID and $\kappa$ is a dimensionless parameter given by $\kappa^2 = \sqrt{E_C/E_L}/2$.

The tunneling of Cooper pairs in the spectrometer  is associated with the absorption of photons by the rf-SQUID~\cite{griesmar2021superconducting}. Therefore, the resulting dc-current $I_J$ flowing in the spectrometer is proportional to the photon absorption rate $\Lambda$, $I_J = 2e\Lambda$. Treating the coupling Hamiltonian Eq.~\eqref{eq:HamiltonianCoupling} as a time-dependent perturbation to an unperturbed Hamiltonian Eq.~\eqref{eq:HamiltonianRF},  $\Lambda$ can be calculated using the Fermi's golden rule~\cite{Note1}
\begin{align}
\Lambda =  \dfrac{2\pi}{\hbar}\left(2 E_LA(\varphi_s)\right)^22\kappa^2\Big|\langle i|\hat{X}| f\rangle\Big|^2 \rho(E_f),
\label{eq:Lambda}
\end{align}
where $\langle i |\hat{X} | f \rangle$ is the matrix element calculated between the initial and final states of $H_{rf}$, and $\rho(E_f)$ is the density of states at the energy $E_f$ of the final states. 

However, in the absence of the dissipation, the photon absorption rate will have delta-peaks when the excitation energies of $H_{rf}$ are in resonance with $\omega_J$. To include bath degrees of freedom in our treatment, we formulate the problem in terms of the Keldysh action~\cite{kamenev2011nonequilibrium} and derive the semiclassical equation of motion for the classical coordinate $X_{cl}$~\cite{Note1}. Introducing a new variable $\tilde{X}_{cl}(t)=X_{cl}(t) - \varphi_x/\kappa$, the equation of motion for the classical field $\tilde{X}_{cl}(t)$ reads 
\begin{align}
&\ddot{\tilde{X}}_{cl}(\tau) +  \dfrac{\gamma}{\hbar} \dot{\tilde{X}}_{cl}(t) +  \tilde{X}_{cl}(t)+ 2 \kappa\dfrac{E_J}{\hbar\omega_p} \sin{[\kappa \tilde{X}_{cl}(\tau) +\varphi_x]}  \notag\\
&  =  \dfrac{A(\varphi_s)}{\kappa} \cos{\left(\dfrac{\omega_J}{\omega_p}\tau\right)},
	\label{eq:EQforXcl}
	\end{align}
where $\gamma$ is the dissipation, $\tau = t \omega_p$ is a dimensionless time, and $\omega_p = 1/\sqrt{LC}$.
Expanding $\sin{[\kappa X_{cl}(\tau)]} $ up to third order in $\kappa \ll 1$, Eq.~\eqref{eq:EQforXcl} takes the form of a \emph{generalized Duffing equation}
\begin{align}
&\ddot{\tilde{X}}_{cl}(\tau) +  \dfrac{\gamma}{\hbar} \dot{\tilde{X}}_{cl}(\tau) + \Omega(\varphi_x)\tilde{X}_{cl}(\tau) +
\frac{\partial V_{nl}}{\partial \tilde{X}_{cl}}=f(\tau),
\label{eq:EqforXclphix}
\end{align}
where $\Omega(\varphi_x)$ is the renormalised plasma frequency  given by
\begin{align}
\Omega(\varphi_x) = \omega_p \sqrt{1+ \beta_L\cos{( \varphi_x)}},
\label{eq:Omega}
\end{align}
with $\beta_L = E_J/\left(2E_L\right)$, while $V_{nl}$ accounts for the non-linearity arising from the Josephson energy
$$
V_{nl}=\lambda(\varphi_x)\tilde{X}_{cl}^3(\tau)+\lambda'(\varphi_x) \tilde{X}_{cl}^4(\tau)\,.
$$
We note that the shape of the non-linear potential is fully tunable by $\varphi_x$, since we have  $\lambda(\varphi_x)$$=$$-\kappa^3 E_J\sin{( \varphi_x)}/\left(3\hbar\omega_p\right) $  and
$\lambda'(\varphi_x)$$=$$-\kappa^4 E_J\cos{( \varphi_x)} /\left(12\hbar\omega_p\right)$. Finally, $f(\tau)$ in Eq.~\eqref{eq:EQforXcl}  is the time-dependent drive
$$
f(\tau)=- 2 \kappa\dfrac{E_J}{\hbar\omega_p}\sin{( \varphi_x)}   
+\dfrac{A(\varphi_s)}{\kappa} \cos{\left(\dfrac{\omega_J}{\omega_p}\tau\right)}.
$$
Eq.~\eqref{eq:EQforXcl} describes therefore a nonlinear differential equation in presence of drive and dissipation, whose solution we will discuss in the following.
\begin{figure}[t!] 
	\centering
	\includegraphics[width=0.9\linewidth]{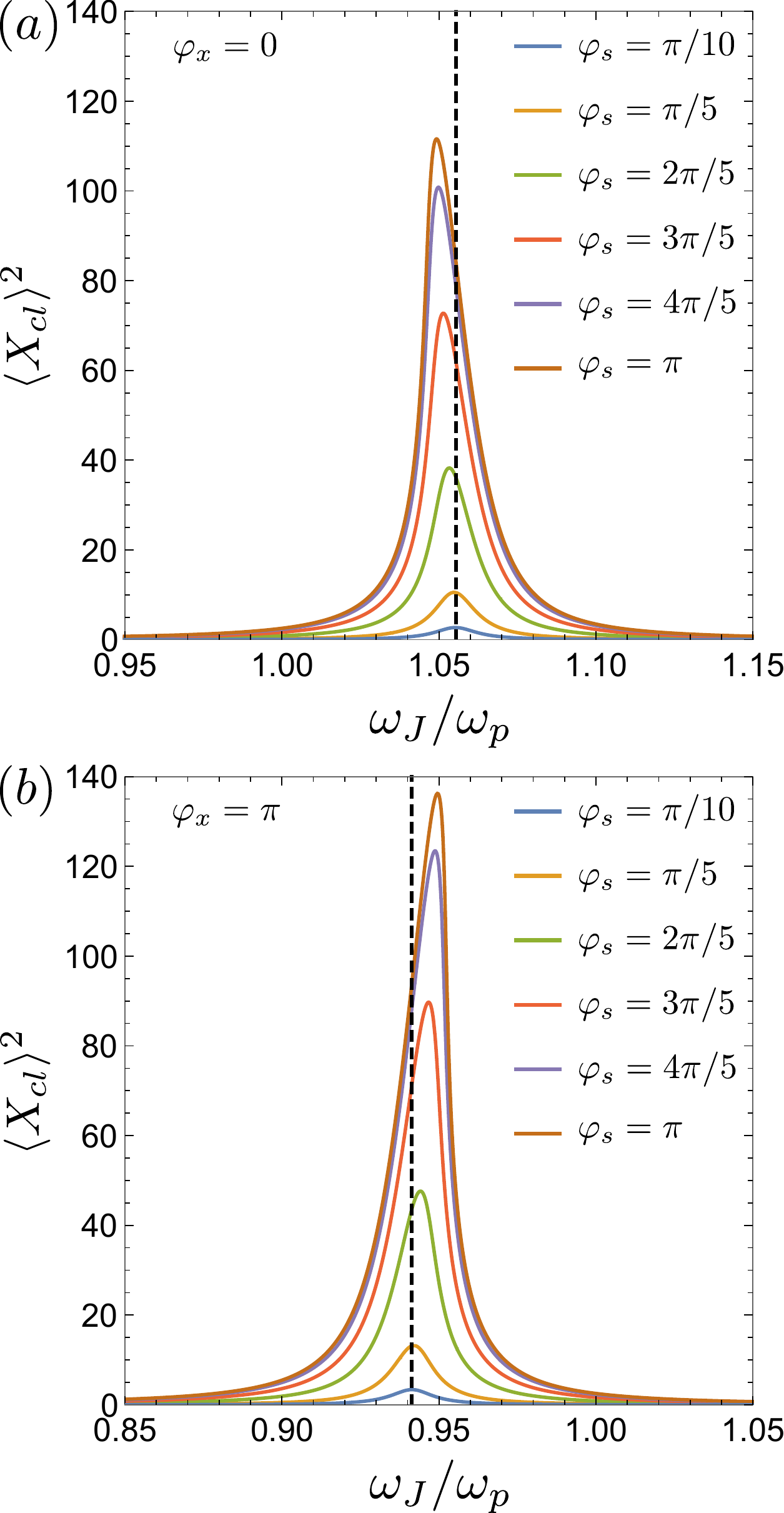}
	\caption{Average value of the coordinate squared $\langle X_{cl}\rangle^2$ as a function of the frequency $\omega_J/\omega_p$ for the phase of the rf-SQUID $\varphi_x = 0$ (top panel) and $\varphi_x = \pi$ (bottom panel).  Different colors of the lines correspond to different values  of $\varphi_s$, from bottom to top: $\varphi_s = \pi/10$, $\varphi_s = \pi/5$, $\varphi_s = 2\pi/5$, $\varphi_s = 3\pi/5$, $\varphi_s = 4\pi/5$, $\varphi_s = \pi$. Black dashed line corresponds to the frequency $\Omega(\varphi_x)$ given by Eq.~\eqref{eq:Omega}. 
Both for (a) $\varphi_x = 0$ and (b) $\varphi_x = \pi$ the average value of the coordinate increases when increasing $\varphi_s$. (a) For $\varphi_x = 0$ the position of the peak in $\langle X_{cl}\rangle^2$  shifts to smaller value of $\omega_J/\omega_p$, resulting in backward-bending of $\langle X_{cl}\rangle^2$.
(b) For $\varphi_x = \pi$ the position of the peak in $\langle X_{cl}\rangle^2$  shifts to larger value of $\omega_J/\omega_p$, resulting in forward-bending of the coordinate squared. Other  parameters are fixed as $\omega_p =2\pi\times  45,91$~GHz, $\beta_L = E_J/\left(2E_L\right) = 0.114$, $L=58$~pH, $L_S=43.7$~pH, $I_{0S}/I_0=1/3$, $k=0.5$, and $\gamma/\hbar = 0.017$.
}
	\label{fig:PhiX=0} 
\end{figure}

{\it Forward-Backward Transition in the non-linear spectroscopy regime.}~---~ 
 Next, we calculate the average value of the coordinate $\langle X_{cl}\rangle$, obtained from the steady-state solution of Eq.~\eqref{eq:EqforXclphix},  for different values of $\varphi_x$ and $\varphi_s$.  While a full numerical solution of the Duffing equation is reported for completeness in Ref.~\cite{Note1}, here we discuss the results using a semi-analytical approach that captures perfectly the features contained in the full numerics.

In absence of any non-linearity the solution of Eq.~\eqref{eq:EqforXclphix} takes the form  
\begin{equation}
X_{cl}(t)=X_{cl}(\omega)\cos(\omega t+\phi),
\end{equation}
where the frequency response $X_{cl}(\omega)$ has a peak at the renormalized plasma frequency $\Omega(\varphi_x)$. In presence of non-linear terms an ansatz of this form does not solve the Duffing equation exactly, yet we can still obtain a closed equation for $X_{cl}(\omega)$  by disregarding higher-order harmonics~\cite{Note1}. Solving this equation for different values of $\varphi_x$ and $\varphi_s$ allow us to obtain the result plotted in Fig.~\ref{fig:PhiX=0}, where we show the frequency response for two different values of $\varphi_x=0,\pi$ considered in Ref.~\cite{griesmar2021superconducting}, and for different values of   $\varphi_s$ corresponding to the evolution from the linear to the non-linear spectroscopy regime.
We see that in the linear spectroscopy regime  the frequency response displays a small peak centered around $\Omega(\varphi_x)$, and the shape of the peak does not change as 
$\varphi_x$ is varied. However, upon increasing the strength of the drive, the response become strongly anharmonic with a peak which increases in size and becomes more and more distorted. In particular we see that upon tuning $\varphi_x$  from zero to $\pi$ the frequency response shows a transition from backward to forward bending. This transition can be immediately understood by noticing that in general, the steady-state solution of the Duffing equation is sensitive to the sign of the coefficient in front of the cubic term. In our case this coefficient depends explicitly from the phase of the rf-SQUID and in particular changes sign at $\varphi_x=\pi/2$
$$
\lambda'(\varphi_x)=-\kappa^4 E_J\cos{( \varphi_x)} /\left(12\hbar\omega_p\right)
$$
leading therefore to a transition in the shape of the frequency response.  Quite interestingly a qualitatively similar behavior was found in the experimental results of Ref~\cite{griesmar2021superconducting}, in particular, in the non-linear spectroscopy regime. We will go back later on this point to present a quantitative comparison with the experimental data.

\begin{figure}[t!] 
	\centering
	\includegraphics[width=0.8\linewidth]{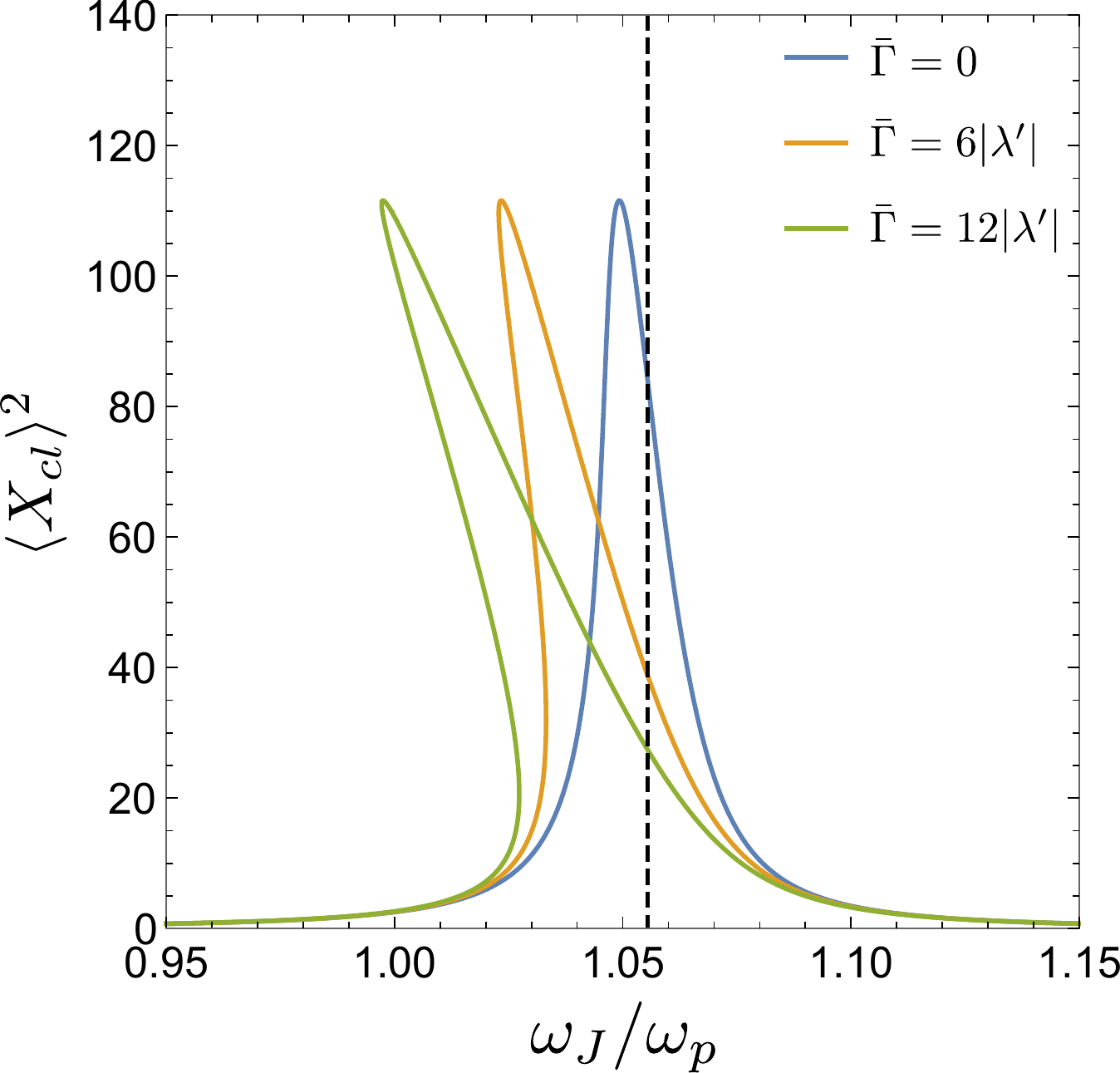}
	\caption{Average value of the coordinate squared $\langle X_{cl}\rangle^2$ as a function of the driving frequency $\omega_J/\omega_p$ in the presence of the feedback from the load line, 
		 $V =  V_J + R\Gamma \langle X_{cl}\rangle^2$. 
		  The phase of the rf-SQUID is fixed to zero, $\varphi_x = 0$. Blue line corresponds to $\bar{\Gamma}=0$, orange line corresponds to $\bar{\Gamma} = 6|\lambda'(0)|$,  and green line corresponds to $\bar{\Gamma} = 12|\lambda'(0)|$. 
	Other parameters are the same as in Fig.~\ref{fig:PhiX=0}.  } 
	\label{fig:NphFeedback} 
\end{figure}

{\it Role of Feedback from Load Line.}~---~ 
 \label{VoltageNumberPhotons}
The previous section have highlighted the role of the Duffing non-linearity and its tunability with the phase $\varphi_x$ at the origin of the forward-backward transition in the frequency response of the rf-SQUID. Here we discuss another source of non-linear behavior, that is at play in the experimental setting of Ref.~\cite{griesmar2021superconducting}, namely the fact that the bias voltage $V$  is not constant, but depends on the current itself as $V = V_J +R I_J$, where $R$ is the resistance in series with the spectrometer. Within our model this implies that the frequency at which the Duffing oscillator is driven, i.e. $\omega_J$ in Eq.~\eqref{eq:EQforXcl}, depends self-consistently on the average value of the  oscillator coordinate,
\begin{equation}
\omega_J\rightarrow \omega_J + \bar{\Gamma}\omega_p\langle X_{cl}\rangle^2
\end{equation}
where we used the relation $I_J = \Gamma \langle X_{cl}\rangle^2$. Here, $\bar\Gamma = 2 e R\Gamma/\left(\hbar\omega_p\right) $. This feedback mechanism introduces an additional source of non-linear behavior as we show in Fig.~\ref{fig:NphFeedback}, where we plot the frequency response for  $\varphi_x = 0$, for fixed value of the Duffing non-linearity, and different values of the feedback parameter $\bar{\Gamma}$.
\begin{figure}[t!] 
	\centering
	\includegraphics[width=\linewidth]{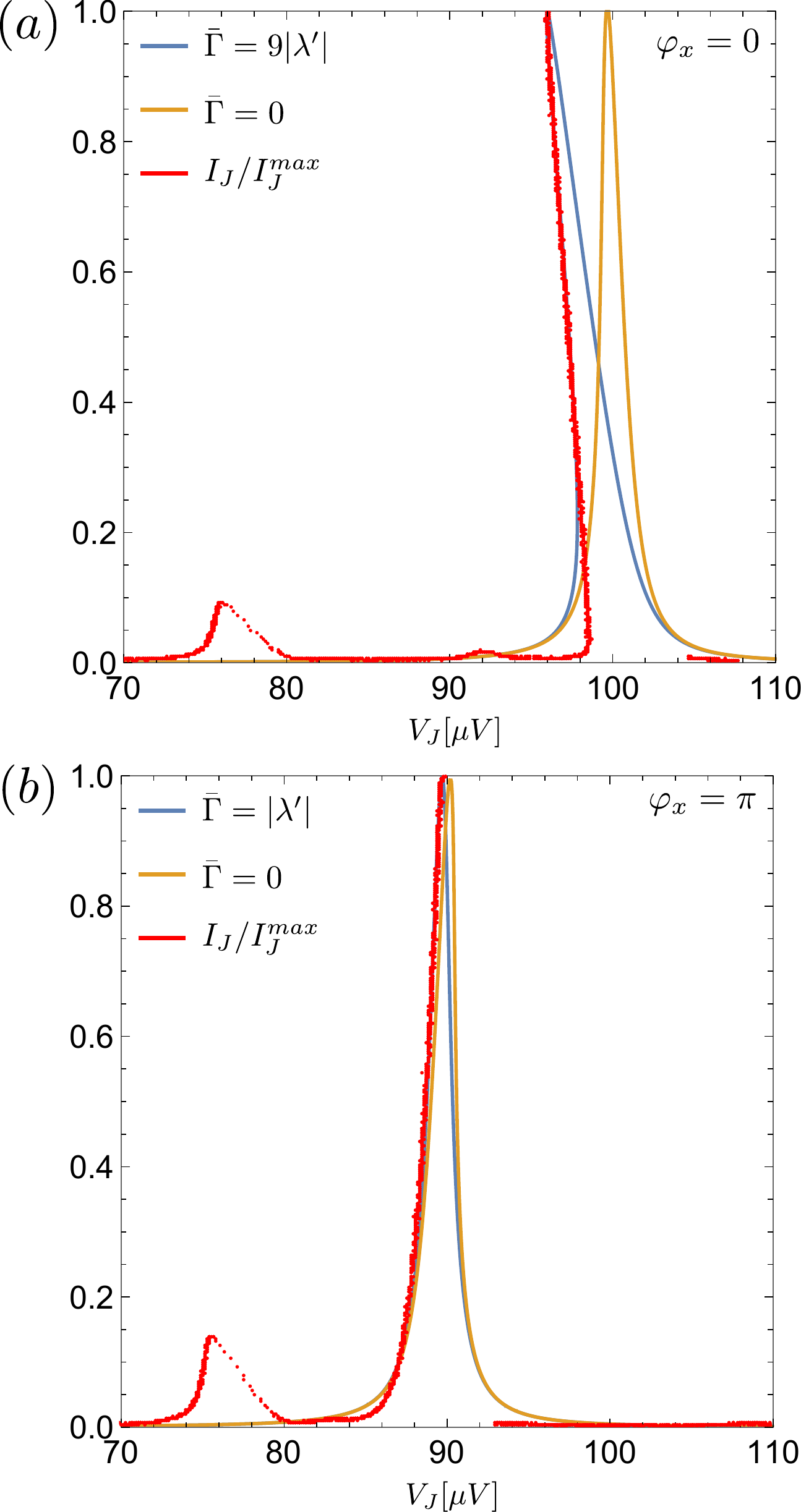}
	\caption{Normalized average value of the coordinate squared $\langle X_{cl}\rangle^2/\langle X_{cl}\rangle^2_{max}$   as a function of bias voltage $V_J$ [$\mu$V] for (a) $\varphi_x = 0$ and (b) $\varphi_s = \pi$. Orange line corresponds to $\langle X_{cl}\rangle^2/\langle X_{cl}\rangle^2_{max}$ calculated  for the constant voltage $V\equiv V_J$  (in the absence of the feedback). Blue line corresponds to $\langle X_{cl}\rangle^2/\langle X_{cl}\rangle^2_{max}$ calculated in the presence of the feedback on the applied voltage, $V =  V_J + R\Gamma \langle X_{cl}\rangle^2$.
	The experimental data for the current (divided by its maximum value) is presented by red dots. 
			A good agreement between $I_J/I_{max}$ and $\langle X_{cl}\rangle^2/\langle X_{cl}\rangle^2_{max}$ (calculated in the presence of the feedback) can be achieved by tuning $\bar{\Gamma} = 9|\lambda'(0)|$ (top) or $\bar{\Gamma} = |\lambda'(\pi)|$ (bottom).
Other parameters are the same as in Fig.~\ref{fig:PhiX=0}.} 
	\label{fig:ComparisonLambda1Phix=0} 
\end{figure}
We see that the strength of the back-bending becomes stronger and stronger upon increasing the feedback effect from the loadline. We further notice that from the experimental parameters used in Ref.~\cite{griesmar2021superconducting} we estimate that the feedback contribution to the voltage is sizeable, of the order of $\delta V_J\sim 38\mu V$ for an averaged measured voltage $V_J\sim 96\mu V$.
In the next section, we are going to present a detailed quantitative comparison with the experimental results that show how both effects, namely tunable Duffing non-linearity and feedback are needed to reproduce the results.

{\it Comparison with Experimental Data.}~---~
We conclude by comparing the prediction of our theory for the frequency response of the rf-SQUID with the experimental data obtained through the Josephson spectrometer. We focus again on two specific values of the rf-SQUID phase,  $\varphi_x = 0,\pi$ showing respectively backward and forward bending and fix the phase of the spectrometer to $\varphi_s=\pi$, namely the strong drive regime. We note that there are two free parameters in the model: $\gamma$ and $\Gamma$, that are not fixed by the measurements~\cite{griesmar2021superconducting}. Let us first consider the case of $\varphi_x = 0$. The current-voltage characteristic  (normalized by its maximum value) is presented in Fig.~\ref{fig:ComparisonLambda1Phix=0} (red line). In the same figure, we plot the average value of the coordinate squared (normalized by its maximum value) in the absence of the feedback, $\bar{\Gamma} = 0$ (orange line). We note that the position of the maximum in current-voltage characteristic in $V_J$ is smaller than the position of the maximum in $\langle X_{cl}\rangle^2$. Therefore, $I_J$ bends stronger than $\langle X_{cl}\rangle^2$ in the absence of the feedback. To find a better agreement between the experimental data and theoretical predictions, we include a feedback effect in the calculation of $\langle X_{cl}\rangle^2$. As expected, the finite $\bar{\Gamma}$ increases the bending of $\langle X_{cl}\rangle^2$ as a function of the bias $V_J$, see Fig.~\ref{fig:ComparisonLambda1Phix=0}~(a) (blue line). Moreover, by choosing a specific value of $\bar{\Gamma}$ we can find a good agreement between experimental data for the current $I_J$ and calculated  $\langle X_{cl}\rangle^2$ in the limit of large number of photons (semiclassical approximation). Similarly, for the case $\varphi_x = \pi$ we see that our theory is able to capture the forward bending but in order to quantitatively reproduce the data the inclusion of the feedback mechanism is important. Furthermore, we note that the agreement with the experimental data is excellent for large values of the current, corresponding to large photon numbers, as expected for our semiclassical theory, while at low intensity quantum fluctuations are likely crucial to capture the sharp edge seen in the current-voltage characteristic.

{\it Conclusions.}~---~ 
We studied an rf-SQUID inductively coupled to the spectrometer based on two Josephson junctions. We calculated the average value of the phase difference across the rf-SQUID, which is proportional to the current flowing in the spectrometer, and found that the position of the peak in the frequency response is given by $\Omega(\varphi_x)$ and, therefore, depends on the rf-SQUID phase $\varphi_x$. For large values of $\varphi_s$, corresponding to the non-linear spectroscopy regime, we found that the peak maximum shifts to higher (lower) values of the frequency for $\varphi_x>\pi/2$ ($\varphi_x<\pi/2$), leading to the forward (backward) bending of the peak. Moreover, taking into account the feedback from the load line allows us to get a quantitative agreement with the experimental data~\cite{griesmar2021superconducting}.

\begin{acknowledgments}
{\it Acknowledgments.}~---~ 
This project has received funding from the European Union’s Horizon 2020 research and innovation programme under the Marie Skłodowska-Curie Grant Agreement No.~892800.
This project has received funding from the  European Research Council (ERC) under the European Union's Horizon 2020 research and innovation programme (grant agreement 636744).
This work was supported by the ANR grant ”NonEQuMat” (ANR-19-CE47-0001). 
The research was also supported by IDEX grant ANR-10-IDEX-0001-02 PSL.
\end{acknowledgments}

\begin{widetext}
	
	\newpage
	\onecolumngrid
	\bigskip 
	
	\begin{center}
		\large{\bf Supplemental Material to `Theory of high-power excitation spectra of rf-SQUID' \\}
	\end{center}
	\begin{center}
		Olesia Dmytruk, R.~H.~Rodriguez, {\c{C}}.~{\"O}.~Girit, and Marco Schir\'o
		\\
		{\it JEIP, USR 3573 CNRS, Coll\`ege de France, PSL Research University, F-75321 Paris, France}
	\end{center}

	In the Supplemental Material, we provide the details on derivation of the model Hamiltonian for an rf-SQUID coupled to a Josephson junction spectrometer and relation between current flowing in the spectrometer and expectation value of phase difference across the rf-SQUID. Moreover, we derive the semiclassical equation of motion for the photonic degrees of freedom within the Keldysh formalism. Also, we present an approximate analytical solution of the Duffing equation for arbitrary values of the rf-SQUID phase, and compare it with the full numerical solution. Finally, we derive the analytical solution for the Duffing equation in the presence of feedback from the load line.
	
	\section{Microscopic Model for rf-SQUID coupled to Josephson Spectrometer} 
	
	\begin{figure}[h!]
		\centering
		\includegraphics[width=0.5\linewidth]{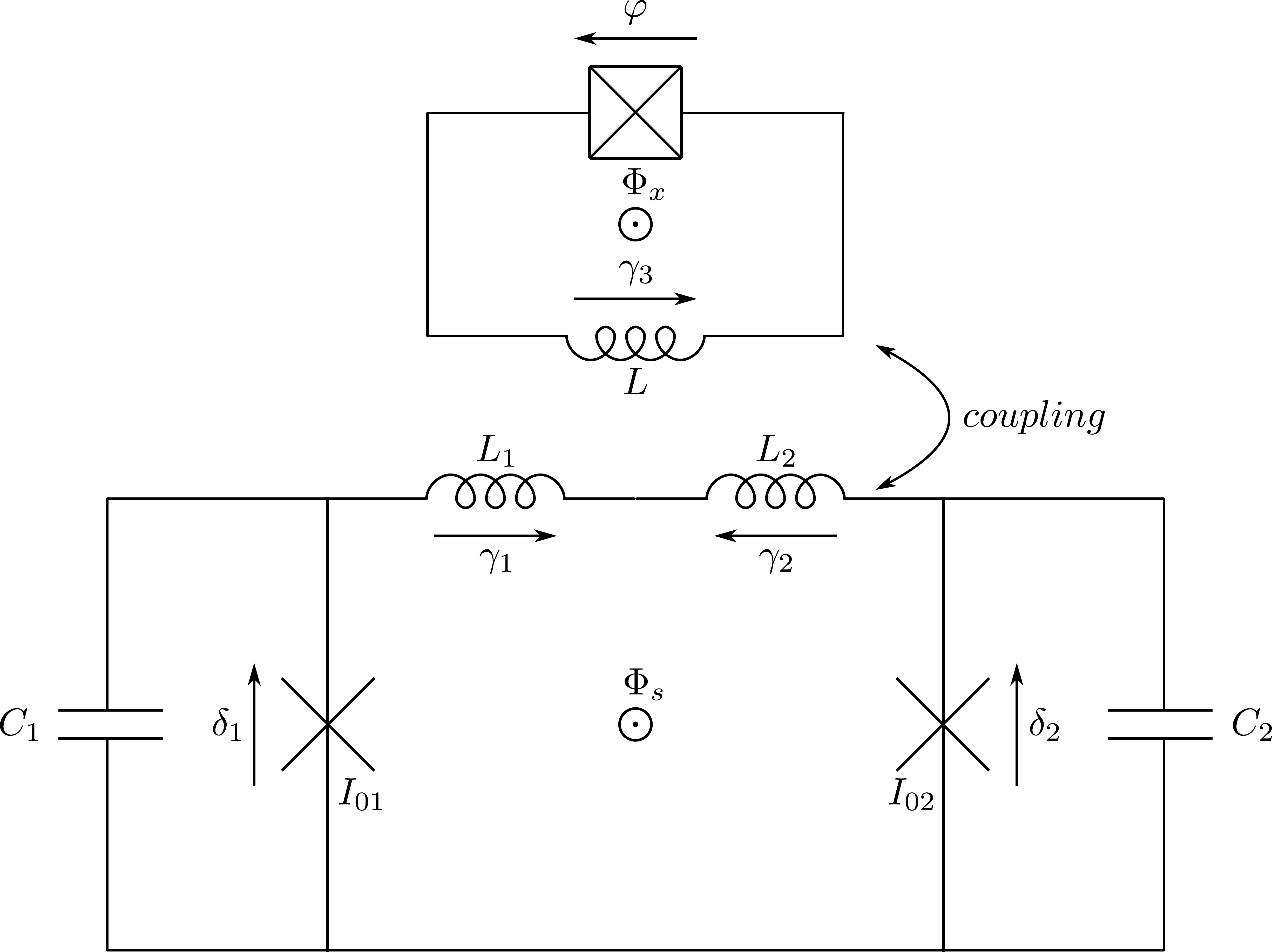}
		\caption{Scheme of the setup: an rf-SQUID (top) coupled inductively to a Josephson junction spectrometer (bottom). The rf-SQUID is formed by a single Josephson junction (boxed cross) embedded in a superconducting loop of inductance $L$. A magnetic flux $\Phi_x$ is threading the loop.
			The Josephson spectrometer consists of two voltage-biased Josephson junctions with capacitance $C_i$ in a superconducting loop of inductance $L_i$, where $i = 1,2$, enclosing a static magnetic flux $\Phi_s$.}
		\label{fig:schemeCouplingNew}
	\end{figure}
	
	In this section, we present detailed derivation of the Hamiltonian that describes an rf-SQUID coupled to a Josephson junction spectrometer presented in the main text. 	First, we write down the Hamiltonian for the rf-SQUID.
	The condition on the phase differences across the rf-SQUID loop reads (see Fig.~\ref{fig:schemeCouplingNew}~(top loop))
	\begin{align}
		\varphi + \gamma_3 = \varphi_x,
	\end{align}
	where $\varphi$ is the phase difference across the junction, $\gamma_3$ is the phase difference across the inductance $L$, and $\varphi_x = \Phi_x/\varphi_0$ is the rf-SQUID phase, with $\Phi_x$ being the magnetic flux threading the  loop and $\varphi_0 = \Phi_0/\left(2\pi\right)$ being the reduced flux quantum.

	The Hamiltonian of the rf-SQUID reads~\cite{makhlin2001quantumSM} 
	\begin{align}
		H_{rf} =  
		E_{C} \hat{N}^2 - E_{J}\cos\hat{\varphi} + E_{L}\left(\hat{\varphi} - \varphi_x\right)^2,
		\label{eq:Hrf}
	\end{align}
	where $E_{C} =2 e^2/C$ is the charging energy, with $C$ being the capacitance of the  junction, $E_{J} = I_{0} \varphi_0$ is the Josephson energy, with $I_{0}$ being the critical current, and $E_{L} = \varphi_0^2/\left(2 L\right)$ is the inductive energy. Here, $\hat{N}$ is the Cooper pairs number operator conjugated to the phase $\hat{\varphi}$, with the commutation relation $[\hat{N},\hat{\varphi}] = -i$.

	Next, we write down the Hamiltonian of the Josephson junction spectrometer.
	The condition on the phase differences across the spectrometer loop is given by (see Fig.~\ref{fig:schemeCouplingNew}~(bottom loop))
	\begin{align}
		\delta_2 + \gamma_2 - \gamma_1 - \delta_1= \varphi_s,
		\label{eq:FluxLoop}
	\end{align}
	where $\delta_{i}$ is the phase difference across the Josephson junction $i = 1,2$ of capacitance $C_{i}$ and critical current $I_{0i}$, $\gamma_{i}$ is the phase difference across the inductance $L_{i}$, and $\varphi_s = \Phi_s/\varphi_0$ is the phase difference across the spectrometer, with $\Phi_s$ being the static magnetic flux threading the spectrometer loop.

	In the limit of small self-inductance $I_{0i} L_i \ll \varphi_0$, the phase $\gamma_i$ can be neglected in Eq.~\eqref{eq:FluxLoop}. Therefore, we obtain that $\varphi_s \approx \delta_2 - \delta_1$.

	The Hamiltonian of the Josephson junction spectrometer can be written as
	\begin{align}
		H_S = E_{C_{J_1}}\hat{N}_{s_1}^2 + E_{C_{J_1}}\hat{N}_{s_1}^2  - E_{J_1}\cos\hat{\delta}_1 - E_{J_2} \cos\hat{\delta}_2 + E_{L_1}\gamma_1^2 + E_{L_2}\gamma_2^2,
		\label{eq:Hspectrometer}
	\end{align}
	where  $E_{C_{J_i}} = 2 e^2/C_i $ is the charging energy of the Josephson junction $i$, $E_{J_i} = I_{0i}\varphi_0$ is the Josephson energy, and $E_{L_i}= \varphi_0^2/\left(2 L_i\right)$ is the inductive energy. Here, $\hat{N}_{si}$ is the Cooper pairs number operator conjugated to the phase $\hat{\delta}_i$, $[\hat{N}_{si},\hat{\delta}_i] = -i$.
	
	Since $\gamma_i \ll1$, the terms in $H_S$ corresponding to the inductive energies can be ignored, while the terms corresponding to the charging energy can be dropped since $E_{C_i}\ll E_{J_i}$.

	Next, we introduce two new variables $\delta$ and $\sigma$ as
	\begin{align}
		&\delta = \dfrac{\delta_2 - \delta_1}{2} \equiv \dfrac{\varphi_s}{2},\\
		&\sigma = \dfrac{\delta_2 +\delta_1}{2},
	\end{align}
	where $\sigma$ is the average phase difference across the spectrometer.

	Let us have a closer look at the term in Eq.~\eqref{eq:Hspectrometer} corresponding to the Josephson energy. Rewriting $\delta_i$ in terms of the new variables and assuming a symmetric SQUID ($L_1 = L_2$, $C_1 = C_2$, $I_{01} = I_{02}$), we arrive at
	\begin{align}
		\cos\delta_1 + \cos\delta_2 = 2\cos\left(\dfrac{\delta_1 - \delta_2}{2}\right)\cos\left(\dfrac{\delta_1 + \delta_2}{2}\right) = 2 \cos\delta\cos\sigma \equiv 2 \cos\left(\varphi_s/2\right)\cos\sigma.
	\end{align}
	Since the power of the spectrometer is maximal at $\varphi_s = \pi$, the Josephson energy in $H_S$ can be also neglected. Therefore, the spectrometer will be included in the total Hamiltonian of the system only through the inductive coupling term.

	Let us now describe the inductive coupling between the spectrometer loop and the rf-SQUID. The spectrometer can only couple with the phase $\hat{\varphi}$, therefore, the coupling Hamiltonian reads
	\begin{align}
		H_{coupl} =  - E_M \left(\gamma_2 - \gamma_1\right) \hat{\varphi},
		\label{eq:Hcoupl}
	\end{align}
	where $E_M = k\varphi_0^2/\sqrt{L L_s} $ is the energy of the mutual inductance between the rf-SQUID and the spectrometer, with $k$ being the dimensionless coupling coefficient, and $L_s$ being the inductance of the symmetric SQUID ($L_1 = L_2 \equiv L_s/2$).

	For a symmetric SQUID, we find that
	\begin{align}
		\gamma_2 - \gamma_1 = \dfrac{I_{0s} L_s }{2\varphi_0}\left(\sin\delta_2 - \sin\delta_1\right) = \dfrac{I_{0s} L_s}{\varphi_0}\sin\left(\varphi_s/2\right)\cos\left(\sigma\right),
		\label{eq:gamma1gamma2}
	\end{align}
	where $I_{0s}$ is the spectrometer critical current. Introducing Eq.~\eqref{eq:gamma1gamma2} into Eq.~\eqref{eq:Hcoupl}, we find that the coupling Hamiltonian  reads
	\begin{align}
		H_{coupl} =  - k\sqrt{\dfrac{L_s}{L}} I_{0s}\varphi_0 \sin\left(\varphi_s/2\right)\cos\left(\omega_J t\right)\hat{\varphi},
		\label{eq:Hcoupl2}
	\end{align}
	where we used that $\sigma = \omega_J t$. Here, $\omega_J = 2 e V_J/\hbar$ is the Josephson frequency, with $V_J$ being the bias voltage.

	Combining Eq.~\eqref{eq:Hrf} and Eq.~\eqref{eq:Hcoupl2}, the total Hamiltonian of the rf-SQUID coupled to the spectrometer reads 
	\begin{align}
		H = H_{rf} + H_{coupl} = E_{C} \hat{N}^2 - E_{J}\cos\hat{\varphi} + E_{L}\left(\hat{\varphi} - \varphi_x\right)^2 - k\sqrt{\dfrac{L_s}{L}} I_{0s}\varphi_0 \sin\left(\varphi_s/2\right)\cos\left(\omega_J t\right)\hat{\varphi}.
		\label{eq:HamiltonianRFS}
	\end{align}
	
	\section{Current flowing through the spectrometer}
	
	In this section, we derive the expression that relates the dc-current $I_J$ and the expectation value of the position operator $\hat{X}$. The dc-current flowing in the spectrometer is proportional to the steady-state photon absorption rate $\Lambda$
	\begin{align}
		I_J = 2 e \Lambda.
		\label{eq:Current}
	\end{align}
	
	Treating the coupling Hamiltonian $H_{coupl}$ as a time-dependent perturbation to an unperturbed Hamiltonian $H_{rf}$, the photon absorption rate $\Lambda$ can be calculated using the Fermi's golden rule
	\begin{align}
		\Lambda = \dfrac{2\pi}{\hbar}\Big|\langle i |H_{coupl} | f \rangle\Big|^2 \rho(E_f),
		\label{eq:Gamma}
	\end{align}
	where $\langle i |H_{coupl} | f \rangle$ is the matrix element of the perturbation $H_{coupl}$ calculated between the initial and final states of $H_{rf}$, and $\rho(E_f)$ is the density of states at the energy $E_f$ of the final states.

	Rewriting Eq.~\eqref{eq:Hcoupl2} in the form
	\begin{align}
		H_k = \tilde{A}(\varphi_s)\cos(\omega_J t)\hat{\varphi},
	\end{align}
	where $\tilde{A}(\varphi_s) =  - k\sqrt{L_s/L} I_{0s}\varphi_0 \sin\left(\varphi_s/2\right)$ and $\hat{X} = \hat{\varphi}/\left(\kappa\sqrt{2}\right)$ is the position operator, we find that the dc-current is given by
	\begin{align}
		I_J = 2e \dfrac{2\pi}{\hbar}\tilde{A}^2(\varphi_s)2\kappa^2\Big|\langle i|\hat{X}| f\rangle\Big|^2 \rho(E_f).
	\end{align}

	Here, $\kappa$ is a dimensionless parameter given by $\kappa^2 = \sqrt{E_C/E_L}/2$. Since the Hamiltonian Eq.~\eqref{eq:HamiltonianRFS} does not include dissipation, the density of states will be a delta-function, with infinite peaks at excitation energies of $H_{rf}$.  Experimentally, dissipation at frequency $\omega_J$ as well as DC voltage fluctuations will result in current peaks of non-zero width. Therefore, it is important to include dissipation in our problem by coupling rf-SQUID modes to a bosonic bath. In the presence of the dissipation, the expectation value of the position operator can be calculated in the semiclassical limit using the Keldysh technique~\cite{kamenev2011nonequilibriumSM}.

	\section{Keldysh action and Semiclassics} \label{KeldyshAction}
	The total action of the system reads
	\begin{align}
		&S = S_{sys} + S_{bath},\\
		&S_{sys}  = S_{osc} + S_{rf} + S_{J}  + S_{c},
	\end{align}
	where
	\begin{align}
		&S_{osc} = \dfrac{1}{2\omega_p}\int_{C} dt \left[\dot{X}(t)^2 - \omega_p^2 X^2(t)\right],\\
		&S_{rf} = \dfrac{2\sqrt{2} E_L \kappa \varphi_x }{\hbar}  \int_{C} dt \ X(t),\\
		&S_{J} = \dfrac{E_J}{\hbar}  \int_{C} dt\ \cos{[\kappa\sqrt{2} X(t)]},\\
		&S_{c} =  2\sqrt{2} E_L  \kappa  A\left(\varphi_s\right) \int_{C} dt\ \cos{(\omega_J t)} X(t),\\
		&S_{bath} =  \int_{C} dt \ \sum_\alpha \dfrac{1}{2\omega_\alpha}\left[( \dot{Y}_\alpha^b)^2 - \omega_\alpha^2 (Y_\alpha^b)^2\right] - \dfrac{2\kappa}{\hbar}\int_{C} dt\  \sum_\alpha g_\alpha  X(t)Y_\alpha^b(t).
	\end{align}
	
	Performing Keldysh rotation for bosonic fields as 
	\begin{align}
		X_{\pm} = \dfrac{1}{\sqrt{2}}(X_{cl} \pm \hbar X_{q}),
	\end{align}
	we arrive at
	\begin{align}
		&S_{osc} = -\dfrac{\hbar}{\omega_p} \int_{-\infty}^{+\infty}dt \ \Big[\ddot{X}_{cl} X_q + \omega_p^2 X_{cl} X_q \Big],\\
		&S_{rf} = 4 E_L \kappa \varphi_x  \int_{-\infty}^{+\infty} dt \ X_q(t),\\
		&S_{J} =  - \dfrac{2E_J}{\hbar} \int_{-\infty}^{+\infty} dt\ \sin{[\kappa X_{cl}(t)]}  \sin{[\kappa \hbar X_{q}(t)]},\label{eq:JosephsonAction}\\
		&S_{c} = 4 E_L \kappa  A\left(\varphi_s\right)  \int_{-\infty}^{+\infty} dt\ \cos{(\omega_J t)} X_q(t),\\
		&S_{bath} = -\int_{-\infty}^{+\infty}dt \ \sum_\alpha \dfrac{\hbar}{\omega_\alpha}\Big[\ddot{Y}^b_{\alpha,cl} Y^b_{\alpha,q}  + \omega_{\alpha}^2 Y^b_{\alpha,cl}Y^b_{\alpha,q} \Big]- 2\kappa \int_{-\infty}^{+\infty} dt\  \sum_\alpha g_\alpha  [X_{cl}(t)Y^b_{\alpha,q}(t) + X_{q}(t)Y^b_{\alpha,cl}(t)].\label{eq:Sbath}
	\end{align}
	
	After integrating out bath degrees of freedom in Eq.~\eqref{eq:Sbath}, we arrive at
	\begin{align}
		S_{bath}[X]
		=  \int_{-\infty}^{+\infty}dt\ dt' \    \Big[X_q(t) g^R_{t,t'}X_{cl}(t')+ X_{cl}(t) g^A_{t,t'} X_q(t') + X_q(t)  g^K_{t,t'}X_q(t')\Big],
	\end{align}
	where we introduced
	\begin{align}
		&g^R_{t,t'} = - \kappa^2 \sum_\alpha  g_\alpha^2 G^R_{\alpha,t,t'},\\
		&\left[G^R_{\alpha,t,t'}\right]^{-1} = -\dfrac{\hbar}{2\omega_\alpha} (\partial_t^2 + (\omega_\alpha)^2)\delta(t-t').
	\end{align}
	
	In what follows, we assume that
	\begin{align}
		g^R_{t,t'} = -\gamma \delta(t-t')\partial_t,
		\label{eq:gRtt'}
	\end{align}
	where $\gamma$ is the dissipation.
	
	If the fluctuations of the quantum component $X_q(t)$ are regarded as small, we can expand $\sin{[\kappa X_{q}(t)]}$ in Eq.~\eqref{eq:JosephsonAction} to the first order and find for the Josephson action
	\begin{align}
		S_{J} \approx - 2E_J \kappa \int_{-\infty}^{+\infty} dt\ \sin{[\kappa X_{cl}(t)]}  X_{q}(t).
		\label{eq:JosephsonAction2}
	\end{align}
	Next, we find that the classical saddle point equation reads [the one that takes $X_q(t) = 0$]
	\begin{align}
		\dfrac{\delta S}{\delta X_q}\Bigg|_{X_q =0} = -\dfrac{\hbar}{\omega_p}\Big[\ddot{X}_{cl}(t)  + \omega_p^2 X_{cl}(t)\Big]+ 4 E_L\kappa\varphi_x  - 2E_J \kappa \sin{[\kappa X_{cl}(t)]} + 4 E_L\kappa A\left(\varphi_s\right) \cos{(\omega_J t)} +  \int dt' \ g^R_{t,t'} X_{cl}(t') = 0.
		\label{eq:classicalsaddlepoint}
	\end{align}
	
	Introducing Eq.~\eqref{eq:gRtt'} into Eq.~\eqref{eq:classicalsaddlepoint}, we arrive at
	\begin{align}
		-\dfrac{\hbar}{\omega_p}\Big[\ddot{X}_{cl}(t)  + \omega_p^2 X_{cl}(t)\Big]+ 4 E_L\kappa\varphi_x  - 2E_J \kappa \sin{[\kappa X_{cl}(t)]} + 4 E_L\kappa A\left(\varphi_s\right) \cos{(\omega_J t)}- \gamma \dot{X}_{cl}(t) = 0,
		\label{eq:classicalsaddlepoint2}
	\end{align}
	that corresponds to the generalized Duffing equation from the main text.

	\section{General Solution for Duffing Equation}
	In this section, we present the details of the derivation of the approximate analytical solution of  Eq.~\eqref{eq:classicalsaddlepoint2} that is used in the main text. First, we introduce a new variable $\tilde{X}_{cl}$ 
	\begin{align}
		\tilde{X}_{cl}(\tau) = X_{cl}(\tau) - \dfrac{\varphi_x}{\kappa}
	\end{align}
	in Eq.~\eqref{eq:classicalsaddlepoint2} and expand the non-linear terms up to third order in $\kappa$, arriving at

	\begin{align}
		&\ddot{\tilde{X}}_{cl}(\tau) +  \dfrac{\gamma}{\hbar} \dot{\tilde{X}}_{cl}(\tau) + \left[1 + 2 \kappa^2\dfrac{E_J}{\hbar\omega_p} \cos{( \varphi_x)} \right] \tilde{X}_{cl}(\tau)   -  \kappa^3\dfrac{E_J}{\hbar\omega_p}\sin{( \varphi_x)}  \tilde{X}_{cl}^2(\tau)-\notag\\
		&-\dfrac{2 \kappa^4} {3!} \dfrac{E_J}{\hbar\omega_p}\cos{( \varphi_x)} \tilde{X}_{cl}^3(\tau) + 2 \kappa\dfrac{E_J}{\hbar\omega_p}\sin{( \varphi_x)} =   4 \kappa \dfrac{E_L}{\hbar\omega_p} A(\varphi_s) \cos{\left(\dfrac{\omega_J}{\omega_p}\tau\right)}.
		\label{eq:EqforXclphixN}
	\end{align}
	
	Next, we rewrite Eq.~\eqref{eq:EqforXclphixN} in the form

	\begin{align}
		\ddot{x} + \alpha \dot{x}+\beta x +\delta x^2 + \zeta x^3 + \eta= \epsilon \cos(\omega t),
		\label{eq:generalizedDuffingEq}
	\end{align}
	
	where we introduce new variables as

	\begin{align}
		&\alpha =\dfrac{\gamma}{\hbar},\\
		&\beta = 1 + 2 \kappa^2\dfrac{E_J}{\hbar\omega_p} \cos{( \varphi_x)},\\
		&\delta = -  \kappa^3\dfrac{E_J}{\hbar\omega_p}\sin{( \varphi_x)} ,\\
		&\zeta = -\dfrac{2 \kappa^4} {3!} \dfrac{E_J}{\hbar\omega_p}\cos{( \varphi_x)} ,\\
		&\eta = 2 \kappa\dfrac{E_J}{\hbar\omega_p}\sin{( \varphi_x)} ,\\
		&\epsilon = 4 \kappa \dfrac{E_L}{\hbar\omega_p} A(\varphi_s) ,\\
		&\omega = \dfrac{\omega_J}{\omega_p}.
	\end{align}

	We look for the solution of Eq.~\eqref{eq:generalizedDuffingEq} in the from
	\begin{align}
		x(t) = a \cos(\omega t) + b\sin(\omega t) + B.
	\end{align}
	Upon neglecting higher order harmonics, we find that the coefficients $a$, $b$ and $B$ are given by a system of equations

	\begin{equation}
		\begin{cases}
			\beta B +\delta \left(\dfrac{a^2}{2}+\dfrac{b^2}{2}+B^2\right) +\zeta \left(\dfrac{3 a^2B}{2} + \dfrac{3b^2 B}{2}+B^3\right) +\eta=0\\
			-a\omega^2 + \alpha b \omega + \beta a + 2 a B \delta +\zeta \left(\dfrac{3 a^3}{4} + \dfrac{3ab^2 }{4}+3 aB^2\right) -\epsilon=0\\
			-b\omega^2 - \alpha a \omega + \beta b + 2 b B \delta +\zeta \left(\dfrac{3 a^2 b}{4} + \dfrac{3b^3 }{4}+3 bB^2\right) =0
		\end{cases}.
		\label{eq:solutionabritraryphix}
	\end{equation}
	
	\begin{figure}[h!] 
		\centering
		\includegraphics[width=0.45\linewidth]{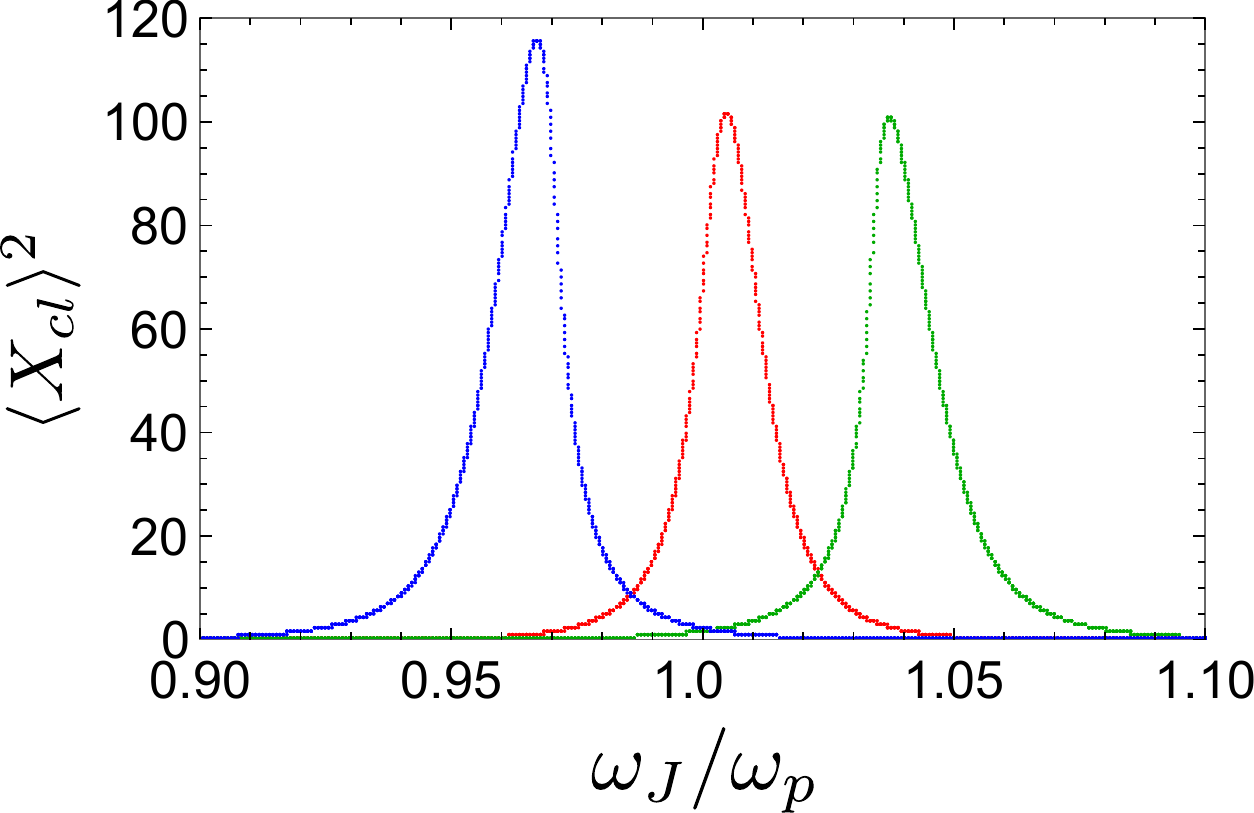}
		\caption{
		Average value of the coordinate squared $\langle X_{cl}\rangle^2$ calculated by numerically integrating Eq.~\eqref{eq:solutionabritraryphix} as a function of the driving frequency $\omega_J/\omega_p$ for different  phases of the rf-SQUID $\varphi_x$. Blue dots correspond to $\varphi_x = 3\pi/4$, green dots correspond to $\varphi_x = \pi/4$, and red dots correspond to $\varphi_x = \pi/2$, signaling the transition  from backward to forward bending of $\langle X_{cl}\rangle^2$. Other  parameters are fixed as $\omega_p =2\pi\times  45,91$~GHz, $\beta_L = E_J/\left(2E_L\right) = 0.114$, $L=58$~pH, $L_S=43.7$~pH, $I_{0S}/I_0=1/3$, $k=0.5$,  $\gamma/\hbar = 0.017$, and $\varphi_s = \pi$.} 
		\label{fig:PhiXPi4} 
	\end{figure}

	Solving Eq.~\eqref{eq:solutionabritraryphix} numerically for $\varphi_s = \pi$, we find the expectation value of $x$ for different values of the rf-SQUID phase. 
	We obtain that for $\varphi_x<\pi/2$, the expectation value of coordinate squared $\langle X_{cl}\rangle^2$ bends backward as a function of the driving frequency, see Fig.~\ref{fig:PhiXPi4}. As the rf-SQUID phase is increased further, $\varphi_x>\pi/2$, $\langle X_{cl}\rangle^2$ bends forward as a function of $\omega_J/\omega_p$, with $\varphi_x = \pi/2$  corresponding to the transition from backward to forward bending.
	Similar dependence of the current-voltage characteristic on $\varphi_x$ was observed in Ref.~\onlinecite{griesmar2021SM}.
	
	Next, we consider two specific cases $\varphi_x = 0$  and $\varphi_x = \pi$ that correspond to $\delta=\eta = 0$ in Eq.~\eqref{eq:generalizedDuffingEq}. Under this choice of the  rf-SQUID phase, Eq.~\eqref{eq:solutionabritraryphix} is reduced to
	
	\begin{equation}
		\begin{cases}
			-a\omega^2 + \alpha b \omega + \beta a  +\zeta \left(\dfrac{3 a^3}{4} + \dfrac{3ab^2 }{4}\right) -\epsilon=0\\
			-b\omega^2 - \alpha a \omega + \beta b  +\zeta \left(\dfrac{3 a^2 b}{4} + \dfrac{3b^3 }{4}\right) =0
		\end{cases},
		\label{eq:solutionabritraryphix2}
	\end{equation}
	and the steady-state solution $z = \sqrt{a^2 + b^2}$  of the Duffing equation can be found from the equation
	
	\begin{align}
		\left((\omega^2 - \beta - 3/4 \zeta z^2)^2 + \alpha^2\omega^2\right) z^2 =\epsilon^2.
		\label{eq:z0}
	\end{align}

	\section{Comparison with Full Numerical Solution of Semiclassics} \label{NumericsPhotonNumber}
	
	In this section, we compare the exact steady-solution for $X_{cl}$ obtained by numerically integrating Eq.~\eqref{eq:classicalsaddlepoint2} and approximated analytical solution given by Eq.~\eqref{eq:z0}. We present the results of such comparison in Fig.~\ref{fig:XclNumerics} for two specific values of $\varphi_x$.
	
	\begin{figure}[h!] 
		\centering
		\includegraphics[width=\linewidth]{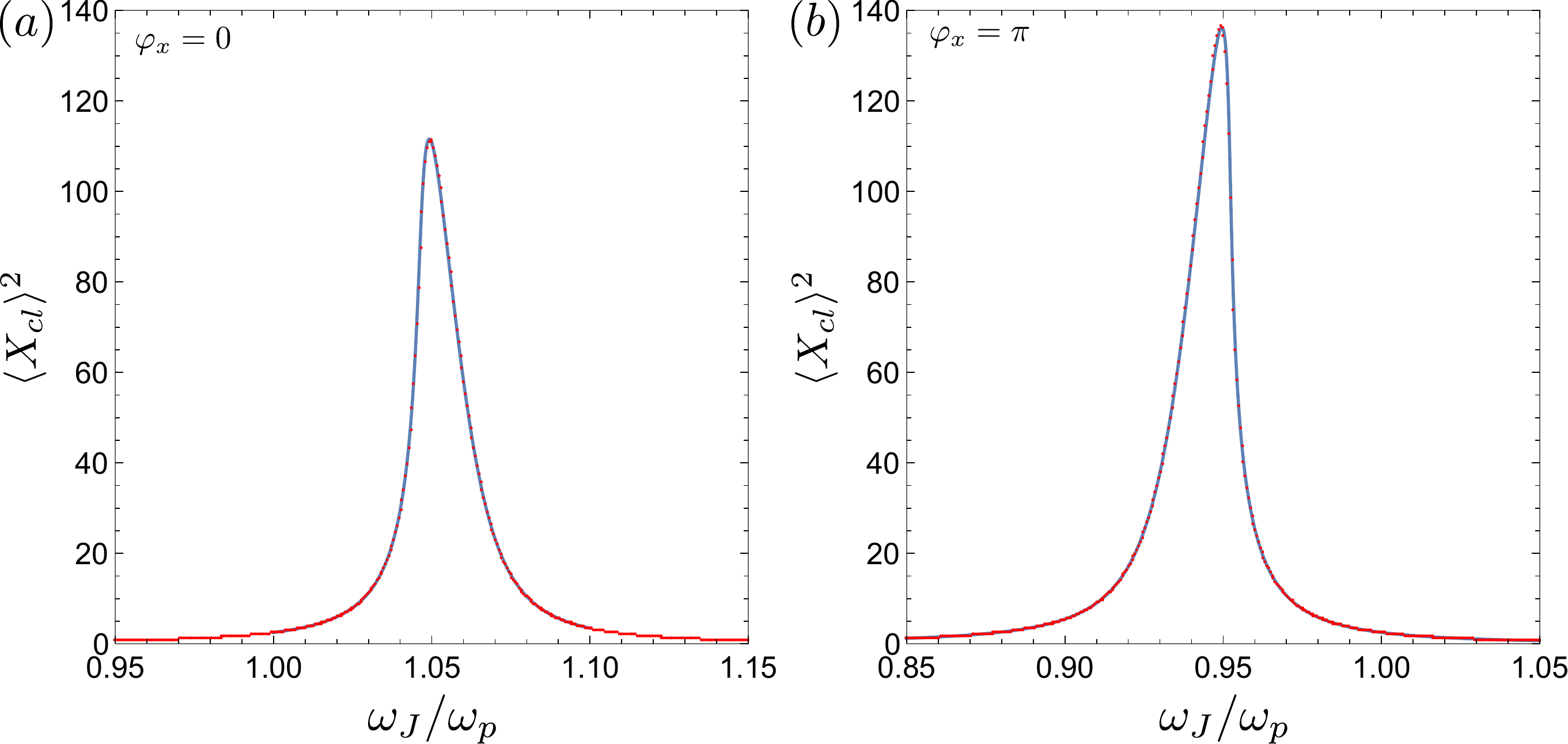}
		\caption{ $\langle X_{cl}\rangle^2$ as a function of the driving frequency $\omega_J/\omega_p$ for (a) $\varphi_x = 0$ and (b) $\varphi_x = \pi$.  Blue solid line corresponds to the approximated analytical solution of the Duffing oscillator, Eq.~\eqref{eq:z0}.    Red dots correspond to the stable solution  calculated by numerically integrating Eq.~\eqref{eq:classicalsaddlepoint2}. There is a good agreement between the full numerical solution and approximated analytical expression for $\langle X_{cl}\rangle$. The parameters are the same as in Fig.~\ref{fig:PhiXPi4} .}
		\label{fig:XclNumerics} 
	\end{figure}

	\section{Solution for Duffing Equation in Presence of Feedback from Load Line}
	The feedback effect from the load line on $\langle X_{cl}\rangle$ can be taken into account by considering $\omega\rightarrow \omega - f z^2$, where  $f = -2 e R \Gamma /\hbar$. In the specific case of $\varphi_x = 0$ or $\varphi_x = \pi$, the solution $z$ can be found by solving the modified Eq.~\eqref{eq:generalizedDuffingEq},
	
	\begin{align}
		\ddot{x} + \alpha \dot{x}+\beta x +\zeta x^3= \epsilon\cos[\left(\omega - f z^2\right) t].
		\label{eq:DuffingOsc4}
	\end{align}
	The coefficients in Eq.~\eqref{eq:DuffingOsc4} and the parameters of our system are related as 
	
	\begin{align}
		&\alpha= \gamma/\hbar,\\
		&\beta = 1 \pm 2 \kappa^2E_J/\left(\hbar\omega_p\right),\\
		&\zeta = \mp \kappa^4 E_J/\left(3\hbar\omega_p\right),\\
		&\epsilon = 4 \kappa E_LA(\varphi_s)/\left(\hbar\omega_p\right) ,\\
		&\omega = \omega_J/\omega_p,\\
		&f = -2 e R \Gamma /\left(\hbar\omega_p\right).
	\end{align}
	Here, the upper (lower) sign in $\beta$ and $\zeta$ correspond to $\varphi_x = 0$ ($\varphi_x = \pi$), while the sign of $f$ remains fixed. Looking for the solution of Eq.~\eqref{eq:DuffingOsc4} in the form $x(t) = z \cos[(\omega - f z^2) t -\phi]$ and neglecting higher order harmonics, we find that the solution of the Duffing equation in the presence of the feedback is given by the equation
	
	\begin{align}
		z^2\left[\left(\beta-(\omega-fz^2)^2 +\dfrac{3\zeta z^2}{4}\right)^2 + \alpha^2(\omega-fz^2)^2\right]^2 = \epsilon^2.
		\label{eq:responseDuffingSelf}
	\end{align}
	
	Therefore, based on the general form of Eq.~\eqref{eq:responseDuffingSelf}, one expects the increase (decrease) of the backward (forward) bending. This can be explained by the fact that
	both $\zeta$ and $f$ have the negative sign for $\varphi_x = 0$, while $\zeta>0$ and $f<0$ for $\varphi_x = \pi$.

\end{widetext}

\end{document}